\newcommand{\bea}{\begin{eqnarray}}
\newcommand{\eea}{\end{eqnarray}}
\newcommand{\be}{\begin{equation}}
\newcommand{\ee}{\end{equation}}
\newcommand{\ben}{\begin{enumerate}}
\newcommand{\een}{\end{enumerate}}
\newcommand{\bi}{\begin{itemize}}
\newcommand{\ei}{\end{itemize}}
\newcommand{\bmi}[1]{\begin{minipage}{#1 cm}}
\newcommand{\emi}{\end{minipage}}

\def\C{\mathcal C}
\def\R{\mathcal R}

\newcommand{\rund}[1]{\left(#1\right)}
\newcommand{\vc}[1]{\mbox{\boldmath $#1$}}
\renewcommand{\d}{{\rm d}}

\newcommand{\eck}[1]{\left[ #1 \right]}
\newcommand{\ave}[1]{\left\langle #1 \right\rangle}
\newcommand{\vt}{\vartheta}
\newcommand{\vp}{\varphi}

\def\llabel#1{\label{sc:#1}}
\def\elabel#1{\label{eq:#1}}

{\catcode`\@=11
\gdef\SchlangeUnter#1#2{\lower2pt\vbox{\baselineskip 0pt \lineskip0pt
  \ialign{$\m@th#1\hfil##\hfil$\crcr#2\crcr\sim\crcr}}}
}

\documentclass[]{aa}
\usepackage{graphicx}
\usepackage{epsfig}
\begin{document}
   \title{The ring statistics -- how to separate E- and
   B-modes of cosmic shear correlation functions on a finite interval}

   \author{
  Peter Schneider
          \inst{1}
          \and
          Martin Kilbinger  \inst{1}
          }

   \offprints{P. Schneider}
\institute{Argelander-Institut f\"ur Astronomie%
  \thanks{Founded by merging
    of the Sternwarte, Radio\-astro\-nomisches Institut and Institut f\"ur
    Astrophysik und Extraterrestrische Forschung der Universit\"at
    Bonn},
  Universit\"at Bonn, Auf dem H\"ugel 71,
  D-53121 Bonn, Germany\\
              \email{peter, kilbinge@astro.uni-bonn.de}
}
             
\titlerunning{The ring statistics for cosmic shear}

   \date{Received ; accepted }

   \abstract{}{Cosmic shear, the distortion of images of distant
   sources by the tidal gravitational field of the large-scale matter
   distribution in the Universe, is one of the most powerful
   cosmological probes. The measured shear field may be due not only
   to the gravitational lensing effect, but may contain systematic
   effects from the measurement process or intrinsic alignment of
   galaxy shapes. One of the main probes for these systematics are the
   division of the shear field into E- and B-mode shear, where lensing
   only produces the former. As shown in a recent note, all currently
   used E-/B-mode separation methods for the shear correlation
   functions $\xi_\pm$ require them to be measured to arbitrarily small and/or
   large separations which is of course not feasible in practice.}
   {We derive second-order shear statistics which provide a clean
   separation into E- and B-modes from measurements of $\xi_\pm(\vt)$
   over a finite interval only. We call these new statistics the
   circle and ring statistics, respectively; the latter is obtained by
   an integral over the former. The mathematical properties of these new
   shear statistics are obtained, as well as specific expressions for
   applying them to observed data.}
   {It is shown that an E-/B-mode separation can be performed on
   measurements of $\xi_\pm$ over a finite interval in angular
   separation, using the ring statistics. We furthermore generalize
   this result to derive the most general class of second-order shear
   statistics which provide a separation of E- and B-mode shear on a
   given angular interval $\vt_{\rm min}\le \vt\le \vt_{\rm max}$. In
   view of these generalization, we discuss the aperture dispersion
   and the shear dispersion and their relation to the shear
   correlation functions. Our results will be of practical use
   particularly for future cosmic shear surveys where highly precise
   measurements of the shear will become available and where control
   of systematics will be mandatory.}{}

   \maketitle
%

\section{\llabel{1}Introduction}
The weak gravitational lensing effect by the large-scale matter
distribution in the Universe is recognized as one of the most powerful
cosmological probes (see Mellier 1999: Bartelmann \& Schneider 2001; 
Refregier 2003; Schneider 2006 for reviews).
First observed in 2000 (Bacon et al. 2000; Kaiser
et al. 2000; van Waerbeke et al. 2000; Wittman et al. 2000), this
cosmic shear effect has advanced tremendously over the past few years
(see, e.g., van Waerbeke et al. 2001, 2002, 2005; Maoli et al.\ 2001;
Jarvis et al.\ 2003, 2005; Hoekstra et al.\ 2002, 2005; Bacon et al.\
2003; Semboloni et al.\ 2005), and large ongoing and planned surveys
will turn cosmic shear into an important tool for cosmology.

In order for satisfying its promises, systematic errors in cosmic
shear measurements must be carefully controlled. Quite a few of those
possible systematics have been identified, including the uncertainties
in the overall shear calibration resulting from the correction of
image smearing by seeing (e.g., Huterer at al.\ 2006) and
uncertainties in the redshift distribution of source galaxies (e.g.,
Ma et al.\ 2006).

One of the important checks for possible systematic effects consists
in investigating the measured shear for B-modes (Crittenden et al.\
2002; hereafter C02), and most of the more recent cosmic shear studies
perform such an analysis. Gravitational lensing effects cause a shear
field which, to leading order, only contains E-modes. The fact that
early cosmic shear measurements contained a significant B-mode
contribution of the shear was interpreted as remaining systematics. As
has been pointed out by several groups (e.g., Heavens et al.\ 2000,
Crittenden et al.\ 2001; Croft \& Metzler 2000; Catelan et al.\ 2000;
Jing 2002), B-modes in the observed shear field can be caused
by intrinsic alignments of source galaxies. This effect can be
controlled in principle if sufficiently accurate photometric redshift
information about the source galaxies are available (King \& Schneider
2002, 2003; Heymans \& Heavens 2003), but this is not the case for
presently available surveys. In addition, B-modes are caused by
gravitational lensing due to the redshift clustering of source
galaxies, but the resulting signal is very small (Schneider et al.\
2002; hereafter S02). If there are inhomogeneous systematic effects,
such as a spatially dependent shear calibration error, B-modes would
also be generated (Guzik \& Bernstein 2005).

The separation of the shear field into E- and B-modes is usually
performed at the level of second-order statistics. Two standard
methods for this have been established in the literature. One of them
is the aperture dispersion statistics (Schneider et al.\ 1998) which
is most conveniently expressed as an integral over the shear two-point
correlation functions $\xi_\pm$ (see C02; S02).  The second method is
the calculation of shear correlation functions which contain only E-
or B-modes (C02, S02).  As was recently shown by Kilbinger et al.\
(2006), these two methods suffer from the fundamental problem that in
order to apply them, one would in principle need to measure the shear
correlation functions to zero separation (in the case of the aperture
dispersion) or to infinite separation (for the E/B-mode
correlation). Both of these measurements are impossible, as pairs of
galaxy images blend if they have too low an angular separation, and
the finite size of observational fields prevents measurements at
arbitrary separations.  This problem of principle becomes a problem in
practice if one studies the presence of B-modes on small and large
angular scales.

In this paper, we present methods for overcoming this problem of
principle, by defining integrals over the shear correlation functions
which (i) extend over the finite range $\vt_{\rm min}\le\vt\le\vt_{\rm
max}$ where the range of integration can be chosen arbitrarily, and
which (ii) provide a clean separation of the shear signal into E- and
B-modes. The first set of such integrals is termed the circle
statistics; it is defined as the correlation of the mean tangential
and cross-component of the shear on two circles with different
radius. As we will show in Sect.\ts\ref{sc:2}, this correlation can be
written in terms of integrals over the shear correlation
function. Since the mean tangential (cross) shear on a circle depends
solely on the E- (B-) mode of the shear field, this provides a clean
separation of both modes. In Sect.\ts\ref{sc:3} we consider the
correlation of the mean tangential and cross shear over two rings
which can as well be written in terms of integrals over the shear
correlation functions $\xi_\pm$. The relation of these new statistics
to the aperture mass is briefly discussed in Sect.\ts\ref{sc:4}. We
then present a generalization of the circle and ring statistics in
Sect.\ts\ref{sc:5} which allows the choice of very general weighted
integrals over the shear correlation functions that provide clear
separation into E- and B-modes. We briefly conclude in
Sect.\ts\ref{sc:6}.

\section{\llabel{2}The circle statistics}
\subsection{\llabel{2.1}Definition}
Consider a circle of radius $\theta$ around the origin and define the
mean of the tangential and cross component of the shear on this
circle, $\C_{\rm t}$ and $\C_\times$, respectively,
measured with respect to the center, 
\bea
\C(\theta)&=&\C_{\rm t}(\theta)+{\rm i}\C_\times(\theta) \nonumber \\
&=&
{1\over 2\pi}\int_0^{2 \pi}\d\vp\;
\rund{\gamma_{\rm t}+{\rm i}\gamma_\times}(\theta,\vp)\;.
\elabel{1}
\eea
The tangential and cross-components of the shear are related to the
Cartesian shear components $\gamma^{\rm c}=\gamma_1+{\rm i}\gamma_2$
through a rotation, 
\be
\gamma_{\rm t}+{\rm i}\gamma_\times=-\gamma^{\rm c}\,{\rm e}^{-2{\rm
    i}\vp}\;,
\elabel{2}
\ee
where $\vp$ describes the polar angle on the circle. As is well known
(see, e.g., C02, S02), $\C_{\rm t}$ is sensitive only to the E-mode of
the shear field, whereas $\C_\times$ is sensitive only to the B-mode.

We next consider the correlation between the mean shear on two
concentric circles with radii $\theta_1$ and $\theta_2$,
\bea
\ave{\C(\theta_1)\C(\theta_2)}
&=&\int_0^{2 \pi}{\d\vp_1\over 2\pi}
\int_0^{2 \pi}{\d\vp_2\over 2\pi}\;
{\rm e}^{-2{\rm i}(\vp_1+\vp_2)}\nonumber \\ &\times&
\ave{\gamma^{\rm c}(\theta_1,\vp_1)\,\gamma^{\rm c}(\theta_2,\vp_2)}
\;,
\elabel{3}
\eea
where we made use of (\ref{eq:2}). Correspondingly, we define the
correlator 
\bea
\ave{\C(\theta_1)\C^*(\theta_2)}
&=&\int_0^{2 \pi}{\d\vp_1\over 2\pi}
\int_0^{2 \pi}{\d\vp_2\over 2\pi}
{\rm e}^{-2{\rm i}(\vp_1-\vp_2)}\nonumber \\ &\times&
\ave{\gamma^{\rm c}(\theta_1,\vp_1)\,\gamma^{\rm c*}(\theta_2,\vp_2)}
\;,
\elabel{4}
\eea
where the asterisk denotes complex conjugation. Expanding the
correlators then yields
\be
\ave{\C(\theta_1)\C(\theta_2)}=\ave{\C_{\rm t}(\theta_1)\C_{\rm t}(\theta_2)}
-\ave{\C_{\times}(\theta_1)\C_{\times}(\theta_2)}\;,
\ee
and the imaginary part vanishes if we assume that the shear field is
parity invariant (Schneider 2003); this assumption is equivalent to
the requirement that the mixed shear correlator $\ave{\gamma_{\rm
t}\gamma_\times}\equiv 0$. Similarly,
\be
\ave{\C(\theta_1)\C^*(\theta_2)}=\ave{\C_{\rm t}(\theta_1)\C_{\rm t}(\theta_2)}
+\ave{\C_{\times}(\theta_1)\C_{\times}(\theta_2)}\;,
\ee
so that the separation into E- and B-modes is achieved with
\be
\ave{\C_{\rm t}(\theta_1)\C_{\rm t}(\theta_2)}
={1\over
  2}\eck{\ave{\C(\theta_1)\C^*(\theta_2)}+\ave{\C(\theta_1)\C(\theta_2)}} \;,
\ee
\be
\ave{\C_{\times}(\theta_1)\C_{\times}(\theta_2)}
={1\over
  2}\eck{\ave{\C(\theta_1)\C^*(\theta_2)}-\ave{\C(\theta_1)\C(\theta_2)}} \;.
\ee

\subsection{\llabel{2.2}Relation to the shear correlation functions} 
The shear correlators that occur in (\ref{eq:3}) and
(\ref{eq:4}) can be expressed in terms of the usual shear correlation
function $\xi_\pm$ by rotating the Cartesian shear into tangential and
cross components relative to the separation direction of the two
points with polar coordinates $(\theta_1,\vp_1)$ and
$(\theta_2,\vp_2)$. We then find that
\bea
\ave{\gamma^{\rm c}(\theta_1,\vp_1)\,\gamma^{\rm c}(\theta_2,\vp_2)}
&=&{\rm e}^{4{\rm i}\vp}\,\xi_-(\vt)\;, \\
\ave{\gamma^{\rm c}(\theta_1,\vp_1)\,\gamma^{\rm c*}(\theta_2,\vp_2)}
&=&\xi_+(\vt)\;,
\eea
where $\vt$ and $\vp$ are the polar coordinates of the separation
vector,
\be
\vt\,{\rm e}^{{\rm i}\vp}
=\theta_2\,{\rm e}^{{\rm i}\vp_2}
-\theta_1\,{\rm e}^{{\rm i}\vp_1}
\;.
\ee
More explicitly, the separation is
\be
\vt^2=\theta_1^2+\theta_2^2-2\theta_1\theta_2\cos(\Delta\vp)\;,
\elabel{12}
\ee
where $\Delta\vp=\vp_2-\vp_1$ is the difference of the polar angle of
the two points under consideration. In the calculation of (\ref{eq:3})
we need the square of the expression
\be
{\rm e}^{2{\rm i}\vp}\,{\rm e}^{-{\rm i}(\vp_1+\vp_2)}
= {
\theta_1^2\,{\rm e}^{-{\rm i}\Delta\vp} +
\theta_2^2\,{\rm e}^{{\rm i}\Delta\vp} - 2\theta_1\theta_2
\over \vt^2} \;.
\ee
The integrand in (\ref{eq:3}) then depends only on the difference
$\Delta\vp$ of the polar angles, so that one of the angular integrals
can be carried out. Furthermore, we note that the integral is real,
since the imaginary part of the integrand is an odd function of
$\Delta\vp$. This then allows us to write the correlator of the circle
statistics as 
\bea
\ave{\C(\theta_1)\C(\theta_2)}
\!\!&=&\!\!\!\int_0^{ \pi}\!{\d\Delta\vp\over \pi}\,{\xi_-(\vt) \over \vt^4}
\,\bigl[
(\theta_1^4+\theta_2^4)\cos(2\Delta\vp) \nonumber \\
\!\!&-&\!\! 4\theta_1\theta_2(\theta_1^2+\theta_2^2)\cos(\Delta\vp)
+ 6 \theta_1^2 \theta_2^2  \bigr]
\;.
\eea
We can now transform the integral into one over the separation $\vt$. 
Using (\ref{eq:12}) we find for the transformation of variables
\be
\d\Delta\vp={2\vt\,\d\vt\over \sqrt{(\theta_1+\theta_2)^2-\vt^2}
\sqrt{\vt^2-(\theta_2-\theta_1)^2}} \;.
\ee
The $\vt$-integration then extends from $\theta_2-\theta_1$ to
$\theta_1+\theta_2$, where we set $\theta_1\le\theta_2$ without loss
of generality. Then, the final expression can be written
as
\be
\ave{\C(\theta_1)\C(\theta_2)}
=\int_{\theta_2-\theta_1}^{\theta_1+\theta_2}
{\d\vt\over \vt}\; \xi_-(\vt) \,
Y_-\rund{{\vt\over\theta_2},{\theta_1\over\theta_2}}\;,
\ee
where the dimensionless function $Y_-$ is
\bea
 Y_-(x,\eta)&=& {1\over \pi\eta^2
 x^2\sqrt{(1+\eta)^2-x^2}\sqrt{x^2-(1-\eta)^2}}
\nonumber \\
&\times& \bigr[(1-x^2)^2-2\eta^2(2-x^2)+\eta^4(6+2x^2+x^4)\nonumber \\
&-& 2\eta^6(2+x^2)+\eta^8 \bigr] \;. 
\eea
In a similar way, we obtain for the other correlator (\ref{eq:4})
\bea
\ave{\C(\theta_1)\C^*(\theta_2)}
\!\!&=&\!\!\!\int_0^{ \pi}\!{\d\Delta\vp\over \pi}\,\xi_+(\vt)
\cos(2\Delta\vp) \nonumber \\
&=&\int_{\theta_2-\theta_1}^{\theta_1+\theta_2}
{\d\vt\over \vt}\; \xi_+(\vt) \,
Y_+\rund{{\vt\over\theta_2},{\theta_1\over\theta_2}}\;,
\eea
with
\be
Y_+(x,\eta)={ x^2\eck{ (1-x^2)^2-2\eta^2 x^2+\eta^4}
\over \eta^2 \pi \sqrt{(1+\eta)^2-x^2}\sqrt{x^2-(1-\eta)^2}} \;.
\ee
We therefore get for the E-mode of the circle statistics
\bea
\ave{\C_{\rm t}(\theta_1)\C_{\rm t}(\theta_2)}
&=&\int_{\theta_2-\theta_1}^{\theta_1+\theta_2}
{\d\vt\over2
\vt}\biggl[\xi_+(\vt)\,Y_+\rund{{\vt\over\theta_2},{\theta_1\over\theta_2}} 
\nonumber \\
&+&\xi_-(\vt) \,Y_-\rund{{\vt\over\theta_2},{\theta_1\over\theta_2}}\biggr]\;,
\elabel{20}
\eea
and the corresponding B-mode reads
\bea
\ave{\C_{\times}(\theta_1)\C_{\times}(\theta_2)}
&=&\int_{\theta_2-\theta_1}^{\theta_1+\theta_2}
{\d\vt\over2
\vt}\biggl[\xi_+(\vt)\,Y_+\rund{{\vt\over\theta_2},{\theta_1\over\theta_2}} 
\nonumber \\
&-&\xi_-(\vt) \,Y_-\rund{{\vt\over\theta_2},{\theta_1\over\theta_2}}\biggr]\;.
\eea
Hence, we have obtained a statistics which is able to perform the
E-/B-mode separation of the shear correlation functions on a finite 
interval in angular separation, $\theta_2-\theta_1 \le \vt\le
\theta_1+\theta_2$. In particular, it avoids the use of correlation
functions towards zero separation which, as we argued before, are
difficult or impossible to be obtained from observations and whose
neglect causes artificial E-/B-mode mixing in the aperture statistics.

However, the circle statistics is not very practical, since the
functions $Y_\pm$ have a $\vt^{-1/2}$ singularity at both integration
boundaries. This singularity has a purely geometrical origin, in that at
the minimum and maximum separation of points on the two circles, the
circles are `mutually parallel', so that pairs with these two extremum
separations are particularly frequent.  Applying the foregoing
equations to observed correlation functions with their noise will
therefore yield estimates for the circle statistics which will be
dominated by the noise just at the boundaries. It it is therefore
desirable to find an expression for the E-/B-mode separation on a
finite interval which does not contain such a singularity. This will
be provided by the ring statistics, which is discussed in
Sect.\ts\ref{sc:3}.

\subsection{\llabel{2.3}Relation to the power spectrum}
\begin{figure}[!tb]
  
  \resizebox{\hsize}{!}{
    \includegraphics{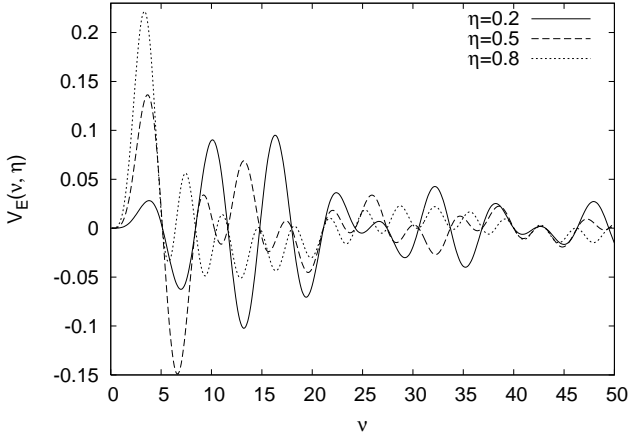}
  }
  
  \caption{The filter function ${\cal V}_{\rm E}$ relating
      the power spectrum to the circle statistics, see Eq.~(\ref{C-P}),
      for three different choices of the ratio of radii $\eta =
      \theta_1/\theta_2$.}
  \label{fig:V}
\end{figure}

If we assume that the shear field is caused by gravitational lensing,
no B-modes are present. The shear field then is a pure E-mode field
which derives from the underlying surface mass density $\kappa$ (we
use standard lensing notation here). The power spectrum $P(\ell)$ of
the $\kappa$ field then specifies the second-order statistics of the
shear field, in particular the correlation functions. We have (see,
e.g., Kaiser 1992)
\be
\xi_\pm(\vt)=\int_0^\infty{\d\ell\;\ell\over 2\pi}\;
P(\ell)\,{\rm J}_{0,4}(\vt\ell)\;,
\ee
where the ${\rm J}_n$ are Bessel functions of the first
kind. Inserting this expression into the E-mode circle statistics
(\ref{eq:20}) yields 
\be
\ave{\C_{\rm t}(\theta_1)\C_{\rm t}(\theta_2)}
=\int_0^\infty{\d\ell\;\ell\over 2\pi}\;P(\ell)\,{\cal
V}_{\rm E}\rund{\ell\theta_2,{\theta_1\over \theta_2}}\;,
\label{C-P}
\ee
where 
\[
{\cal V}_{\rm E}(\nu,\eta)=\int_{1-\eta}^{1+\eta}{\d x\over 2x}\,
\eck{ {\rm J}_0(\nu x)Y_+(x,\eta)+{\rm J}_4(\nu x)Y_-(x,\eta)} \;.
\nonumber
\]
Hence, the function ${\cal V}_{\rm E}(\nu,\eta)$ describes the filter
function with which the circle statistics is related to the lensing
power spectrum; it depends on the ratio $\eta$ of the two radii. The
filter function is plotted in Fig.\ts\ref{fig:V} for three values of
$\eta$. It shows an oscillating behavior which reflects the
oscillations of the Bessel functions. 

The behavior of ${\cal V}_{\rm E}$ for small and large values of $\nu$
can be obtained analytically. Considering small $\nu$ first, we can
expand the Bessel functions in the foregoing equation. The
leading-order term in ${\rm J}_4$ is $(\nu x)^4$, whereas the
expansion of ${\rm J}_0$ contains terms proportional to $(\nu x)^0$
and $(\nu x)^2$. However, we find that
\[
\int_{1-\eta}^{1+\eta}{\d x\over 2x}\;Y_+(x,\eta) =0=\int_{1-\eta}^{1+\eta}{\d
x\over 2x} \,x^2\,Y_+(x,\eta)\;,
\]
so that the leading-order term from the ${\rm J}_0$ part is also
$\propto \nu^4$. Together, we conclude that ${\cal V}_{\rm E}(\nu,\eta)
\propto \nu^4$ for small $\nu$, as can also be seen in
Fig.\ts\ref{fig:V}. 

In the opposite case of large $\nu$, the behavior of the integral for 
${\cal V}_{\rm E}(\nu,\eta)$ is dominated by the
$1/\sqrt{x}$-singularity of the functions $Y_\pm$ at both integration
limits. Hence, the asymptotic behavior of ${\cal V}_{\rm E}(\nu,\eta)$
can be obtained from an integral with very similar properties,
\[
\int_{-x_0}^{x_0}\d x\; { {\rm J}_0(\nu x) \over
\sqrt{x_0^2-x^2} } =
\pi {\rm J}_0^2\rund{\nu x_0\over 2}
\]
and
\[
\int_{-x_0}^{x_0}\d x\; { {\rm J}_4(\nu x) \over
\sqrt{x_0^2-x^2} } =
\pi {\rm J}_2^2\rund{\nu x_0\over 2} \;.
\]
Thus, for large $\nu$ we expect ${\cal V}_{\rm E}(\nu,\eta)$ to be an
oscillating function, where the amplitude of the oscillation
decreases as $\nu^{-1}$. Again, this behavior is seen in
Fig.\ts\ref{fig:V}. 
 
The vanishing of the B-mode implies that
\bea
{\cal V}_{\rm B}(\nu,\eta)&=&\int_{1-\eta}^{1+\eta}{\d x\over 2x}\,
\eck{ {\rm J}_0(\nu x)Y_+(x,\eta)-{\rm J}_4(\nu x)Y_-(x,\eta)} 
\nonumber \\
&\equiv& 0\;
\nonumber
\eea
for all values of $\nu$ and arbitrary $0<\eta<1$. As a consequence, in
the absence of B-modes, the circle statistics can be calculated from
$\xi_+$ and $\xi_-$ separately, since both terms occurring in
(\ref{eq:20}) are then equal.

\section{\llabel{3}The ring statistics}
Instead of the shear on a circle, we consider here the shear inside a
ring and define
\be
\R=\R_{\rm t}+{\rm i}\R_\times
=\int_{\zeta_1}^{\zeta_2}\d\theta\;W(\theta)\,\C(\theta)\;,
\ee
where the integral extends over the annulus $\zeta_1\le \theta\le
\zeta_2$, and $W(\theta)$ is a weight function which we choose to be
normalized to unity, 
\be
\int_{\zeta_1}^{\zeta_2}\d\theta\;W(\theta)=1\;.
\ee
\begin{figure}[!tb]
  
  \resizebox{\hsize}{!}{
    \includegraphics{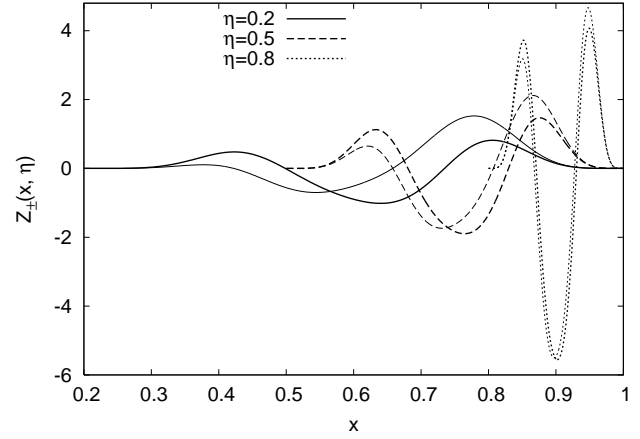}
  }
  
  \caption{The functions $Z_+$ (thick curves) and $Z_-$ (thin curves)
    as functions of the angular separation scaled by the maximum separation,
    $x=\vt/\vt_{\max}$, for three different values of the ratio $\eta
    = \vt_{\rm min}/\vt_{\rm max}$. The specific choice for the ring
    boundaries (\ref{eq:26}) and weight functions (\ref{eq:28}) was assumed.}
  \label{fig:Z}
\end{figure}
As was true for the circle statistics, the real and imaginary parts of
$\R$ are solely due to E-modes and B-modes, respectively. We now
consider the correlator of two such concentric rings, one with
$\zeta_1\le \theta\le 
\zeta_2$ as before, the other with $\zeta_3\le \theta\le
\zeta_4$. If $\zeta_i<\zeta_j$ for all $i<j$, as we shall assume from here
on, the rings are non-overlapping. The maximum and minimum separation
of points from one ring to the other are $\vt_{\rm
max}=\zeta_2+\zeta_4\equiv \Psi$ and $\vt_{\rm
min}=\zeta_3-\zeta_2\equiv\eta\Psi$, respectively. The correlator
reads
\bea
\ave{\R\R}&=&\int_{\zeta_1}^{\zeta_2}\d\theta_1\;W_1(\theta_1)
\int_{\zeta_3}^{\zeta_4}\d\theta_2\;
W_2(\theta_2)\ave{\C(\theta_1)\C(\theta_2)} \nonumber \\
&=&\int_{\zeta_1}^{\zeta_2}\d\theta_1\;W_1(\theta_1)
\int_{\zeta_3}^{\zeta_4}\d\theta_2\;
W_2(\theta_2) 
\elabel{24}
\\
&\times&\int_{\theta_2-\theta_1}^{\theta_1+\theta_2}
{\d\vt\over \vt}\; \xi_-(\vt) \,
Y_-\rund{{\vt\over\theta_2},{\theta_1\over\theta_2}}
\; . \nonumber
\eea
\begin{figure*}[!tb]
  
  \begin{center}
    \resizebox{0.8\hsize}{!}{
      \includegraphics{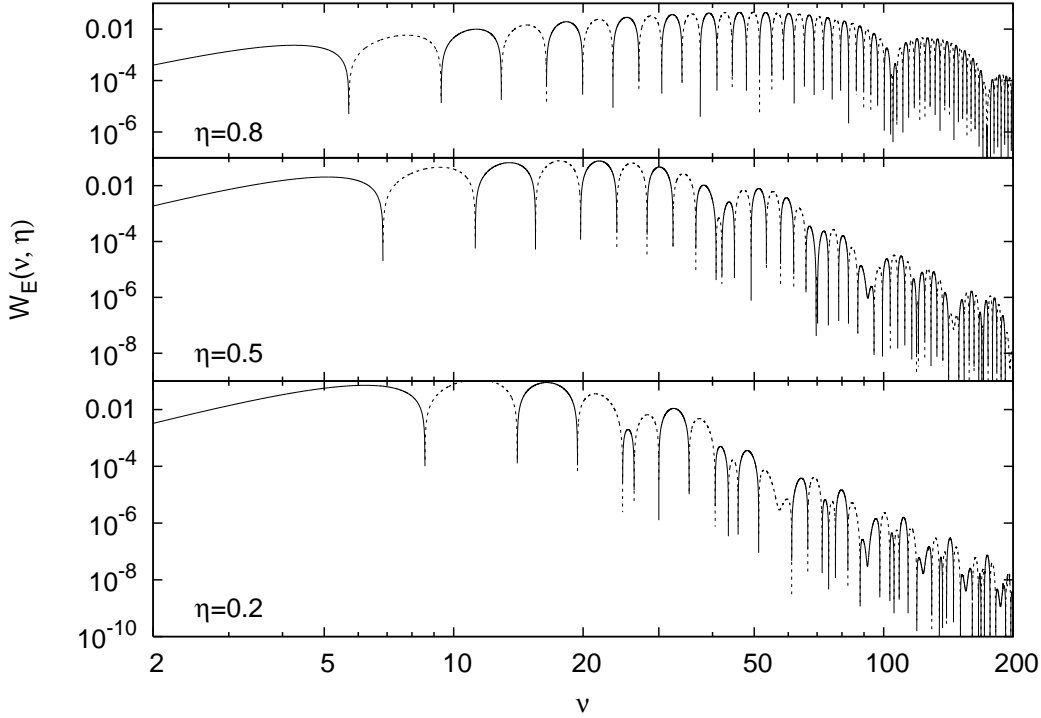}
    }
  \end{center}
  
  \caption{The filter function ${\cal W}_{\rm E}$ relating the power
    spectrum to the ring statistics, see eq.~(\ref{W-P}), for three
    different choices of the ratio of radii $\eta =
    \theta_1/\theta_2$. Dotted curves indicate negative values.}
  \label{fig:W}
\end{figure*}
Interchanging the order of integration, this correlator can be written
in the form 
\be
\ave{\R\R}=\int_\eta^1{\d x\over x}\;\xi_-(x\Psi)\,Z_-(x,\eta)\;,
\elabel{25}
\ee
where $\eta=\vt_{\rm min}/\vt_{\rm max}$ is the ratio of minimum and
maximum separation of points between the two rings, and the function
$Z_-(x,\eta)$ can be calculated explicitly by changing the
integration order in (\ref{eq:24}). In general, the resulting form of
the function $Z_-$ will be complicated, due to the numerous case
distinctions in the range of integrations. The task simplifies
considerably if special ratios of the ring radii are chosen. In the
appendix, we provide explicit formulae for $Z_-$ for the following
choice:
\bea
\zeta_1&=&{1-\eta\over 8}\Psi\; ;\;
\zeta_2={3(1-\eta)\over 8}\Psi\;, \nonumber \\
\zeta_3&=&{5\eta+3\over 8}\Psi\; ;\;
\zeta_4={3\eta+5\over 8}\Psi\;.
\elabel{26}
\eea
This choice still allows the freedom to choose the range
$\eta\Psi\le\vt\le\Psi$ over which the correlation function is probed
in the ring statistics. In analogy, the other correlator involving
$\xi_+$ reads
\be
\ave{\R\R^*}=\int_\eta^1{\d x\over x}\;\xi_+(x\Psi)\,Z_+(x,\eta)\;.
\ee
To be specific, we will choose for the weight functions
\bea
W_1(\theta_1)&=&{30(\theta_1-\zeta_1)^2(\zeta_2-\theta_1)^2
\over(\zeta_2-\zeta_1)^5}\; ,\nonumber \\
W_2(\theta_2)&=&{30(\theta_2-\zeta_3)^2(\zeta_4-\theta_2)^2
\over(\zeta_4-\zeta_3)^5}\;.
\elabel{28}
\eea
In Fig.~\ref{fig:Z} we have plotted the functions $Z_\pm$ for three values of
$\eta$. Both functions $Z_+$ and $Z_-$ have two roots in the interval
$\eta< x<1$, starting off with positive values near the edges of their
range of support. We will explain this qualitative behavior in
Sect.\ts\ref{sc:5.2} below. 

The E- and B-modes of the ring statistics are given by the
combinations
\bea
\ave{\R\R}_{\rm E}&=&{1\over 2}\rund{\ave{\R\R^*} + \ave{\R\R}}\;,
\nonumber \\
&=&\int_\eta^1{\d x\over
2x}\,\eck{\xi_+(x\Psi)\,Z_+(x,\eta)+\xi_-(x\Psi)\,Z_-(x,\eta)} \;,
\nonumber\\
\ave{\R\R}_{\rm B}&=&{1\over 2}\rund{\ave{\R\R^*} - \ave{\R\R}} 
\elabel{31}\\
&=&\int_\eta^1{\d x\over
2x}\,\eck{\xi_+(x\Psi)\,Z_+(x,\eta)-\xi_-(x\Psi)\,Z_-(x,\eta)} \;.
\nonumber
\eea
In case the shear field derives from gravitational lensing, its
B-modes vanish, and the ring statistics $\ave{\R\R}_{\rm E}$ can be
expressed in terms of the power spectrum $P(\ell)$ of the projected
surface mass density. In analogy to the circle statistics, we obtain
\be
\ave{\R\R}_{\rm E}=\int_0^\infty{\d\ell\;\ell\over 2\pi}\;P(\ell)\,{\cal
W}_{\rm E}\rund{\ell\Psi,\eta}\;,
\label{W-P}
\ee
where 
\[
{\cal W}_{\rm E}(\nu,\eta)=\int_\eta^1 {\d x\over 2x}\,
\eck{ {\rm J}_0(\nu x)Z_+(x,\eta)+{\rm J}_4(\nu x)Z_-(x,\eta)} \;.
\]
In Fig.~\ref{fig:W} we have plotted the function ${\cal W}_{\rm
E}(\nu,\eta)$ which is the filter function that relates the power
spectrum to the ring statistics, for three different values of $\eta$.
As we can see, ${\cal W}_{\rm E}$ is a smooth function for small
$\nu$, whereas it strongly oscillates for larger values of $\nu$.
Asymptotically, ${\cal W}_{\rm E}(\nu, \eta)$ goes as $\nu^4$ for
small $\nu$, and the amplitude of oscillations falls off as
$\nu^{-7}$ for large $\nu$.

\begin{figure}[!tb]
  
  \resizebox{\hsize}{!}{
    \includegraphics{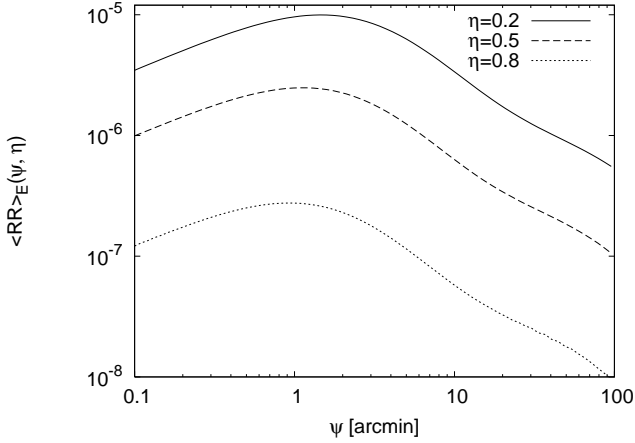}
  }
  
  \caption{The ring statistics $\ave{\R\R}_{\rm E}$ (\ref{W-P}) as a
    function of the maximum separation $\Psi = \vt_{\rm max}$ for
    which the correlation functions need to be known, for three
    different ratios $\eta = \vt_{\rm min}/\vt_{\rm max}$.}
    \label{fig:R}
\end{figure}

The ring statistics $\ave{\R\R}_{\rm E}$ is plotted in
Fig.~\ref{fig:R}, for a $\Lambda$CDM model with $\sigma_8=0.85$ and a
mean redshift of source galaxies of $\bar z \approx 1.5$. For this
calculation, the non-linear power spectrum according to the
prescription of Smith et al.\ (2003) has been used. We see that
$\ave{\R\R}_{\rm E}$ has a very small amplitude which decreases for
larger $\eta$. Hence, the narrower the interval is over which the
correlation functions are considered, the smaller is the measured
shear signal. For all values of $\eta$, the signal of $\ave{\R\R}_{\rm
E}$ is much smaller than the characteristic values of the shear
correlation functions. The mathematical reason for this can be seen in
Fig.\ts\ref{fig:Z} in combination with (\ref{eq:31}): to derive
$\ave{\R\R}_{\rm E}$, the shear correlation is integrated over the
functions $Z_\pm$ which has two roots in their range of
support. Another way to see this is by considering the function ${\cal
W}_{\rm E}$ (Fig.\ts\ref{fig:W}) in connection with (\ref{W-P}). Due
to its oscillating behavior, the resulting integral in (\ref{W-P})
is expected to be small. The narrower the ring is (i.e., the larger
$\eta$), the more oscillations does the function ${\cal W}_{\rm E}$
show near its broad maximum, and the larger is the near-cancellation
of power in the integral (\ref{W-P}).

The vanishing of the B-mode then implies that
\bea
{\cal W}_{\rm B}(\nu,\eta)&=&\int_\eta^1 {\d x\over 2x}\,
\eck{ {\rm J}_0(\nu x)Z_+(x,\eta) - {\rm J}_4(\nu x)Z_-(x,\eta)} 
\nonumber \\
&\equiv& 0
\eea
for all values of $\nu$. We have checked numerically that the
functions $Z_\pm$ indeed satisfy this relation.

\section{\llabel{4}Relation to the aperture statistics}
The definition (24) of the ring statistics is very similar to that of
the aperture mass (Kaiser 1995; Schneider 1996), which is defined as
\be
M_{\rm ap}=\int \d^2\vt\;\kappa(\vc\vt)\,U(|\vc\vt|)\;,
\ee
where $U$ is a radial weight function and the integral extends over
the support of $U$. Provided $U$ is a compensated filter function,
$\int\d\vt\,\vt\,U(\vt)=0$, $M_{\rm ap}$ can be written in terms of the
tangential shear, 
\be
M_{\rm ap}=\int \d^2\vt\;\gamma_{\rm t}(\vc\vt)\,Q(|\vc\vt|)\;,
\elabel{35}
\ee
where the weight function $Q$ is related to $U$ by
\be
Q(\theta)={2\over\theta^2}\int_0^\theta\d\vt\;\vt\,U(\vt) -
U(\theta)\;.
\elabel{36}
\ee
The definition of $\R$ is equivalent to (\ref{eq:35}) in the absence
of B-modes, if we set $W(\theta)=2\pi\theta Q(\theta)$. It is
therefore interesting to see whether the ring statistics can be
expressed also in terms of the surface mass density. This indeed is
the case, since (\ref{eq:36}) can be reversed. Multiplying
(\ref{eq:36}) with $\theta^2$ and differentiating with respect to
$\theta$ yields the differential equation
\be
U'=-Q'-{2\over \theta}Q\;,
\ee
which can be readily integrated, using the boundary condition
$U(\theta)\to 0$ as $\theta\to \infty$:
\be
U(\theta)=\int_\theta^\infty{2\d\vt\over \vt}\;Q(\vt)-Q(\theta)\;.
\ee
It is straightforward to check that $U$ is a compensated filter
function. Furthermore, if $Q(\theta)=0$ for $\theta>\theta_{\rm max}$,
then also $U(\theta)=0$ for $\theta>\theta_{\rm max}$. Therefore,
there is an equivalent $U$-filter with which the ring statistics can
be expressed in terms of the surface mass density; $U$ is a positive
constant for $0\le\theta\le\zeta_1$, then decreases and eventually
becomes negative somewhere in the interval
$\zeta_1<\theta<\zeta_2$, and gets zero again for
$\theta\ge\zeta_2$.

\section{\llabel{5}Generalization}
The relations (\ref{eq:31}) that define integrals over the shear
correlation functions which are sensitive separately to E- and B-modes
are very similar to analogous equations for the aperture dispersions
(see S02, C02). We could easily find other forms of the function
$Z_\pm$ by taking different choices of the weight functions $W_{1,2}$
entering the definition of the ring statistics. 

Therefore, the question arises whether one can find a wide class of
functions, analogous to $Z_\pm$ in (\ref{eq:31}) which provides a
separation of the shear correlation into E- and B-modes. This question
will be answered here.

\subsection{\llabel{5.1}The general E/B-mode decomposition}
Thus, we define second-order shear statistics in the following form:
\bea
E&=&\int_0^\infty\d\vt\;\vt\eck{\xi_+(\vt)T_+(\vt)+\xi_-(\vt)T_-(\vt)} \;,
\nonumber \\
B&=&\int_0^\infty\d\vt\;\vt\eck{\xi_+(\vt)T_+(\vt)-\xi_-(\vt)T_-(\vt)} \;,
\elabel{39}
\eea
and look for pairs of function $T_\pm(\vt)$ for which $E$ contains
only E-modes, and $B$ contains only B-modes. In terms of the E/B-mode
power spectra, the correlation functions read (see S02)
\bea
\xi_+(\vt)&=&\int_0^\infty {\d\ell\;\ell\over 2\pi}\,
{\rm J_0}(\ell\vt)\eck{P_{\rm E}(\ell)+P_{\rm B}(\ell)}\;,
\nonumber \\
\xi_-(\vt)&=&\int_0^\infty {\d\ell\;\ell\over 2\pi}\,
{\rm J_4}(\ell\vt)\eck{P_{\rm E}(\ell)-P_{\rm B}(\ell)}\;.
\elabel{40}
\eea
Inserting (\ref{eq:40}) into (\ref{eq:39}) and interchanging the order
of integration, we obtain
\bea
E&=&\int_0^\infty {\d\ell\;\ell\over 2\pi}\,
\Bigl(  P_{\rm E}(\ell)\eck{t_+(\ell)+t_-(\ell)} 
\nonumber \\
&&\qquad\qquad +P_{\rm B}(\ell)\eck{t_+(\ell)- t_-(\ell)} \Bigr) \;,
\nonumber \\
B&=&\int_0^\infty {\d\ell\;\ell\over 2\pi}\,
\Bigl(  P_{\rm E}(\ell)\eck{t_+(\ell)-t_-(\ell)} 
 \\
&&\qquad\qquad +P_{\rm B}(\ell)\eck{t_+(\ell)+ t_-(\ell)} \Bigr) \;,
\nonumber
\eea 
where we have defined the Hankel transform of the functions $T_\pm$,
\bea
t_+(\ell)&=&\int_0^\infty\d\vt\;\vt\, T_+(\vt)\,{\rm J_0}(\ell\vt)\;,
\nonumber \\
t_-(\ell)&=&\int_0^\infty\d\vt\;\vt\, T_-(\vt)\,{\rm J_4}(\ell\vt)\;.
\elabel{42}
\eea
Thus, the functions $E$ and $B$ contain only E- and
B-modes, respectively, if and only if the two Hankel transforms are
identical, 
\be
t_+(\ell) \equiv t_-(\ell) \;.
\elabel{43}
\ee
Any pair of functions $T_\pm$ which satisfies (\ref{eq:43}) leads to a
separation into E- and B-modes. The functions $Y_\pm$ for the circle
statistics and the functions $Z_\pm$ for the ring statistics are
special examples of this, as are the pairs of functions $T_\pm$ for
the aperture statistics and $S_\pm$ for the shear dispersion, as
defined in S02.

We will show next, by explicit construction, that one can choose one
of the two functions, $T_+$ or $T_-$, arbitrarily and the other
function can be uniquely calculated such that (\ref{eq:43}) is
satisfied. First we assume that the function $T_-$ is chosen; then we
calculate $T_+$ from the inverse Hankel transform,
\[
T_+(\vt)=\int_0^\infty\d\ell\;\ell\,{\rm J_0}(\ell\vt)\,t_+(\vt)\;.
\]
Due to the requirement (\ref{eq:43}) we can replace $t_+$ by $t_-$ and
obtain
\bea
T_+(\vt)&=&\int_0^\infty\d\ell\;\ell\,{\rm J_0}(\ell\vt)
\int_0^\infty \d\theta\;\theta\,{\rm J_4}(\ell\theta)\,T_-(\theta)
\nonumber \\
&=&\int_0^\infty \d\theta\;\theta\,T_-(\theta)
\,G(\vt,\theta)\;,
\elabel{44}
\eea
where we defined the function
\be
G(\vt,\theta)=
\int_0^\infty\d\ell\;\ell\,{\rm J_0}(\ell\vt)\,{\rm
J_4}(\ell\theta)\;.
\ee
In S02 it was shown that
\be
G(\vt,\theta)=\rund{{4\over\theta^2}-{12\vt^2\over \theta^4}}{\rm
H}(\theta-\vt) +{1\over\theta}\delta_{\rm D}(\theta-\vt)\;,
\elabel{46}
\ee
where ${\rm H}(x)$ is the Heaviside step function and $\delta_{\rm D}$
the Dirac delta function. Thus,
\be
T_+(\vt)=T_-(\vt)+\int_\vt^\infty\d\theta\;\theta\,T_-(\theta)
\rund{{4\over\theta^2}-{12\vt^2\over \theta^4}}\;.
\elabel{47}
\ee
Hence, for any function $T_-(\vt)$ we can
calculate from (\ref{eq:47}) the corresponding function $T_+$ 
so that this pair of functions satisfies (\ref{eq:43}). 
In a similar way, $T_-$ can be obtained from a chosen function $T_+$,
from 
\be
T_-(\vt)=T_+(\vt)+\int_0^\vt\d\theta\;\theta\,T_+(\theta)
\rund{ {4\over \vt^2}-{12\theta^2\over \vt^4}}\;.
\elabel{48}
\ee

\subsection{\llabel{5.2}General E/B-mode decomposition on a finite interval}
We now return to the original question of this paper, namely to find a
separation of E- and B-modes from correlation functions given on a
finite interval. Thus we choose a function $T_-(\vt)$ which is
non-zero only on the interval $\vt_{\rm min}\le \vt\le \vt_{\rm max}$. 
The upper limit of the integral in (\ref{eq:47}) can then be replaced
by $\vt_{\rm max}$, and we see immediately that $T_+(\vt)=0$ for 
$\vt>\vt_{\rm max}$. If we require that $T_+(\vt)$ vanishes for
$\vt<\vt_{\rm min}$, then the following two conditions need to be
satisfied:
\be
\int_{\vt_{\rm min}}^{\vt_{\rm max}}{\d\vt\over \vt}\,T_-(\vt)
=0=\int_{\vt_{\rm min}}^{\vt_{\rm max}}{\d\vt\over \vt^3}\,T_-(\vt)\;.
\elabel{49}
\ee
Thus, if we choose a function $T_-$ with is non-zero only on a finite
interval, the corresponding function $T_+$ vanishes outside this
interval as well, provided $T_-$ satisfied the two integral
constraints (\ref{eq:49}). The argument which led to the constraints
(\ref{eq:49}) only applies for $\vt_{\rm min}>0$; for $\vt_{\rm
min}=0$, these conditions become obsolete.

Conversely, if we choose a function $T_+$
which vanishes outside the interval $\vt_{\rm min}\le \vt\le \vt_{\rm
max}$, the corresponding function $T_-$ is zero outside the same
interval if $T_+$ satisfies the two constraints
\be
\int_{\vt_{\rm min}}^{\vt_{\rm max}}\d\vt\, \vt\,T_+(\vt)
=0=\int_{\vt_{\rm min}}^{\vt_{\rm max}}\d\vt\, \vt^3\,T_+(\vt)\;
\elabel{50}
\ee
if $\vt_{\rm max}<\infty$.

In fact, it is straightforward to show that the definition of $T_+$ in
terms of $T_-$ as given by (\ref{eq:47}) implies the relations
(\ref{eq:50}) provided $\vt_{\rm min}>0$; this latter condition is
needed to guarantee the existence of the integrals that
occur. Similarly, the definition (\ref{eq:48}) implies the validity of
(\ref{eq:49}) if $\vt_{\rm max}<\infty$.

We can check these results in a different way, by expanding the
Bessel functions into a power series and comparing the result, when
inserted into (\ref{eq:42}) term-by-term. Using
\bea
{\rm J_0}(z)&=&1-{z^2\over 4}+\rund{z\over 2}^4
\sum_{k=0}^\infty { (-z^2/4)^k\over (k+2)!\,(k+2)!} \;,
\nonumber \\
{\rm J_4}(z)&=&\rund{z\over 2}^4
\sum_{k=0}^\infty { (-z^2/4)^k\over k!\,(k+4)!} \;,
\eea
we see that the integral (\ref{eq:42}) of each term is guaranteed to
exist if $T_\pm$ have finite support, which we assume henceforth.
The first two terms of ${\rm J_0}$, when inserted into the first of
(\ref{eq:42}), vanish because of (\ref{eq:50}). We next compare 
terms proportional to $\ell^{(2k+4)}$ of $t_\pm$. For $t_-$, this term
is
\be
t_{-k}=\int\d\vt\;\vt\,T_-(\vt) 
{(-\vt^2/4)^{(k+2)} \over k!\,(k+4)!} \;,
\ee
whereas for $t_+$, the corresponding term reads
\be
t_{+k}=\int\d\vt\;\vt\,T_+(\vt) {(-\vt^2/4)^{(k+2)} \over
(k+2)!\,(k+2)!} \;.
\ee
Inserting now the expression (\ref{eq:47}) for $T_+$, we find
\bea
t_{+k}&=&{(-4)^{-(k+2)}\over (k+2)!\,(k+2)!} \int_0^\infty \d\vt\;
\vt^{(2k+5)} \nonumber \\
&\times&
\eck{T_-(\vt)+\int_\vt^\infty\d\theta\;\theta\,T_-(\theta)
\rund{{4\over\theta^2}-{12\vt^2\over \theta^4}}}
\eea
For the second term, we interchange the order of integration, after
which the $\vt$-integral can be evaluated. This then leads to 
\bea
t_{+k}&=& \int_0^\infty \d\vt\;
\vt\,{(-\vt^2/4)^{(k+2)} \over (k+2)!\,(k+2)!}
 \nonumber \\
&\times&
T_-(\vt)\eck{1+{2\over k+3}-{6\over k+4}}\;.
\eea
Using the definition of the faculty, it is now easy to show that 
$t_{+k}=t_{-k}$ for all $k$. Hence, for weight functions with finite
support, the equivalence (\ref{eq:43}) can be demonstrated for each
order of $\ell$.

It is interesting to consider two special cases of the definitions $E$
and $B$ that have been investigated in the literature. The first case
is that of the aperture dispersion, as discussed in S02. The
corresponding functions there were also denoted by $T_\pm$,
\bea
&&T_+(x)={6(2-15x^2)\over 5}\eck{1-{2\over\pi}\arcsin\rund{x\over 2}}
\nonumber \\
&&+ {x\sqrt{4-x^2}\over 100\pi}
\rund{120+2320x^2-754x^4+132 x^6-9x^8} 
\elabel{N33}
\eea
for $0\le x \le 2$, and zero otherwise, and
\be
T_-(x)={192\over 35\pi}x^3\rund{1-{x^2\over 4}}^{7/2}\,{\rm H}(2-x)\;.
\elabel{N31}
\ee
Here, $x=\vt/\theta_0$, where $\theta_0$ denotes the aperture radius.
The functions are plotted in S02; they both vanish for $x>2$. If we
consider $T_+$ as given, then $T_-$ follows from (\ref{eq:48}); since
$T_+$ satisfies (\ref{eq:50}), $T_-$ shares the property of $T_+$
to vanish for $x>2$. Conversely, consider $T_-$ as given. Since
$T_-=0$ for $x>2$ one obtains immediately from 
(\ref{eq:47}) that $T_+$ shares this property. However, $T_-$ does not
satisfy the relations (\ref{eq:49}); in fact, since $T_-\propto x$ for 
$x\ll 1$, the second of the integrals in (\ref{eq:49}) does not
exist. 

The second example is that corresponding to the shear dispersion in a
circle, where the corresponding weight functions have been denoted as
$S_\pm$ in S02. They are
\bea
S_+(x)&=&{1\over\pi}\eck{4\arccos\rund{x\over 2}-x\sqrt{4-x^2}}
{\rm H}(2-x) \;,\\
S_-(x)&=&{x\sqrt{4-x^2}(6-x^2)-8(3-x^2)\arcsin(x/2)
\over \pi x^4}
\eea
for $x\le 2$, and $S_-(x)=4(x^2-3)/x^4$ for $x>2$. Thus, whereas $S_+$
is confined to a finite interval $0\le x\le 2$, $S_-$ is not. The
reason for this is that $S_+$ does not satisfy (\ref{eq:50}); in fact,
$S_+(x)$ is a non-negative function.

The constraints (\ref{eq:49}) and (\ref{eq:50}) also explain the
qualitative behavior of the functions $Z_\pm$ shown in
Fig.\ts\ref{fig:Z}. The two constraints in either case require that
the functions $Z_\pm$ have two roots in their range of support. As
this was the reason for the small amplitude of the resulting shear
signal shown in Fig.\ts\ref{fig:R}, it becomes clear that this small
amplitude is not due to the specific choice of our ring statistics,
but is a generic feature of the more general statistic $E$.

\section{\llabel{6}Summary and Conclusions}
In this paper we have constructed new second-order cosmic shear
statistics which can be used to separate E- and B-modes from
measurements of the shear correlation function $\xi_\pm$ on a finite
interval. The circle and ring statistics were constructed, based on
the fact that the cross-component of the shear averaged over a circle
vanishes for pure E-mode shear, and the tangential shear component,
when integrated over a circle, vanishes for a pure B-mode shear. Both
of these statistics are obtained by integrating the shear correlation
functions, multiplied with properly defined weight functions, over a
finite interval in angular separation. By choosing the parameters of
the ring and/or circle, the interval on which the shear correlation
shall be probed can be adjusted.

Motivated by the result, we have shown that there exists a very large
family of such weight functions which can be used to separate E- and
B-modes on the second-order statistics level from measured shear
correlations on a finite interval. Hence, there is a large freedom to
choose these weight functions, of which the circle and ring statistics
are special cases. 

The resulting signal of the E- and B-mode shear statistics is small;
this was explicitly demonstrated for the ring statistics (see
Fig.\ts\ref{fig:R}), but as we argued in Sect.\ts\ref{sc:5.2}, such a
small signal is expected in general and can be traced back to the
integral constraints (\ref{eq:49}) and (\ref{eq:50}) that the weight
function $T_\pm$ have to satisfy if they are to probe the shear
correlation for E-/B-modes over a finite interval in angular
separation. The small signal implies a correspondingly small
signal-to-noise for these statistics. In particular, if fairly narrow
ranges of $\vt_{\rm min}\le \vt\le\vt_{\rm max}$ are to be probed, the
corresponding shear data must have high statistical accuracy. We
therefore expect our results here to be of particular relevance for
the upcoming large cosmic shear surveys.

The ring and circle statistics presented here can be generalized to
higher-order cosmic shear statistics as well. The aperture statistics
that were defined to quantify third-order shear statistics (Jarvis et
al.\ 2004, Schneider et al.\ 2005) are plagued with the same problem
as the second-order aperture statistics pointed out by Kilbinger et
al.\ (2006). Hence, one could generalize the circle statistics to
$\ave{ {\cal C}(\theta_1){\cal C}(\theta_2){\cal C}(\theta_3)}$ and
related expressions (by taking complex conjugates of different
factors), project out the E- and B-modes, and express the result in
terms of the shear three-point correlation functions. By taking
weighted integrals over finite ranges of $\theta_i$, the third-order
ring statistics can be obtained. In this way, also the third-order
shear can be tested for the presence of B-modes.

\begin{acknowledgements}
This work was supported 
by the Deutsche Forschungsgemeinschaft under the project
SCHN 342/6--1 and by the DFG Schwerpunktpgrogramm 1177 `Witnesses of
cosmic history'. 
\end{acknowledgements}

\begin{appendix}
\section{Explicit calculation of the functions $Z_\pm$}
\begin{figure}[!tb]
  
  \begin{center}
    \resizebox{0.7\hsize}{!}{
      \includegraphics{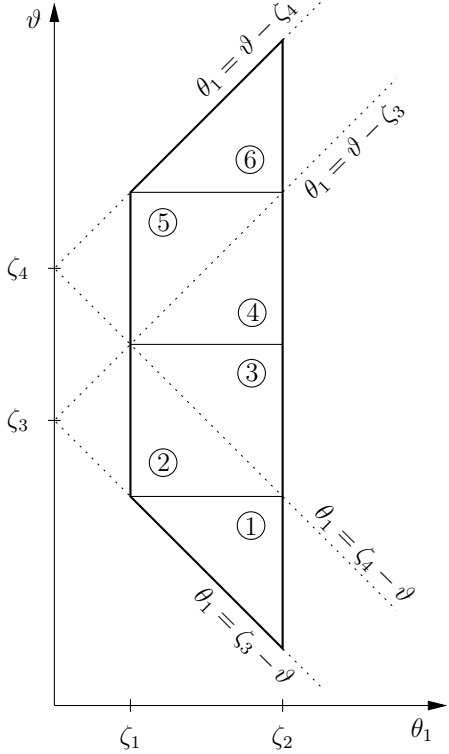}
    }
  \end{center}
  
  \caption{The trapezium (solid line) is the integration area in the
      $\theta_1$-$\vt$-plane for Eq.\ts (\ref{RR}). The devision
      into six triangular regions is indicated, the numbers correspond
      to the indices of the integration bounds in (\ref{RR6}).}
  \label{fig:plane}
\end{figure}

In this appendix, we carry out the steps that lead to the form
of (\ref{eq:25}), and provide explicit expressions for the function
$Z_-$. Starting with the final form of (\ref{eq:24}), then first
interchanging the order of integration between $\theta_2$ and $\vt$,
after that changing the order of integration between $\theta_1$ and
$\vt$, we obtain
\bea
\ave{\R\R}&=&\int_{\zeta_3-\zeta_2}^{\zeta_2+\zeta_4} 
{\d\vt\over \vt}\; \xi_-(\vt)
\int_{{\rm max}(\zeta_1,\vt-\zeta_4,\zeta_3-\vt)}^{\zeta_2}
\d\theta_1\;W_1(\theta_1)\nonumber \\
&\times&\int_{{\rm max}(\zeta_3,\vt-\theta_1)}^{{\rm
min}(\zeta_4,\vt+\theta_1)}
\d\theta_2\;W_2(\theta_2)\,
Y_-\rund{{\vt\over\theta_2},{\theta_1\over\theta_2}}
\; .
\label{RR}
\eea
The integration range in the $\theta_1$-$\vt$ plane is sketched in
Fig.~\ref{fig:plane}. For this figure, we have chosen the values of $\zeta_i$ to
satisfy the relations (\ref{eq:26}); as it becomes clear from this
figure, the integral then naturally splits up into six parts.
Changing to dimensionless variables by defining
\be
\vt=x\Psi\;;\;\theta_1=y_1\Psi\;;\;\theta_2=y_2\Psi\;,
\ee
with $\Psi=\zeta_2+\zeta_4$,
we can then write
\bea
\ave{\R\R}=\sum_{i=1}^6 \!\!\!&&\!\!\!
\int_{a_i}^{A_i}{\d x\over x}\,\xi_-(x\Psi)
\int_{b_i}^{B_i}\d y_1\,w_1(y_1) \nonumber \\
&\times&
\int_{c_i}^{C_i}\d y_2\,w_2(y_2)
Y_-\rund{{x\over y_2},{y_1\over y_2}}\;,
\label{RR6}
\eea
where the various integration bounds are given as
\bea
a_1&=&\eta \; ;\quad A_6=1 \; ;\quad
A_1=a_2=a_3={3\eta+1\over 4}\nonumber\\
a_4&=&a_5=A_2=A_3={1+\eta\over 2}\; ;\quad
a_6=A_4=A_5={\eta+3\over 4}\nonumber
\eea
\bea
b_1&=&-b_4=-B_5={5\eta+3\over 8}-x \nonumber \\
B_1&=&B_3=B_4=B_6={3(1-\eta)\over 8}  \nonumber \\
b_2&=&b_5={1-\eta\over 8}\nonumber \\
B_2&=&b_3=-b_6={3\eta+5\over 8}-x \nonumber 
\eea
\bea
c_1&=&c_2=c_3=c_4={5\eta+3\over 8}\; ;\quad
c_5=c_6=x-y_1 \nonumber \\
C_1&=&C_2=x+y_1\; ;\quad
C_3=C_4=C_5=C_6={3\eta+5\over 8} \;,\nonumber
\eea
and the functions $w_i(y_i)=\Psi W_i(y_i\Psi)$.
For the choice (\ref{eq:28}) of the functions $W_i$, we obtain
\bea
w_1(y_1)&=&{15(8y_1+\eta-1)^2(3-3\eta-8y_1)^2\over 2(1-\eta)^5}\;,\nonumber \\
w_2(y_2)&=&{15(8y_2-5\eta-3)^2(3\eta+5-8y_2)^2\over 2(1-\eta)^5}\;.
\eea
The expressions for $\ave{\R\R^*}$ are identical, except that $\xi_-$
and $Y_-$ are replaced by $\xi_+$ and $Y_+$, respectively.
From these explicit expressions, the functions $Z_\pm(x,\eta)$ can be
read off readily. They are plotted in Fig.~\ref{fig:Z}.

\end{appendix}

\end{document}